\documentclass[amsmath,amssymb,aip,apl,reprint,superscriptaddress]{revtex4-1}

\usepackage{graphicx}
\usepackage{dcolumn}
\usepackage{bm}
\usepackage{siunitx}
\usepackage[colorlinks=true, linkcolor=blue, citecolor=blue, urlcolor=blue]{hyperref}

\def\emph#1{\textcolor{red}{#1}}
\def\emph#1{\textcolor{black}{#1}}

\begin{document}


\title{Roles of heating and helicity in ultrafast all-optical magnetization\\ switching in TbFeCo}


\author{Xianyang Lu}
\affiliation{Department of Physics, University of York, York, YO10 5DD, UK}
\affiliation{York-Nanjing International Joint Center in Spintronics, School of Electronic Science and Engineering, Nanjing University, Nanjing, 210093, China}

\author{Xiao Zou}
\affiliation{Department of Physics, University of York, York, YO10 5DD, UK}

\author{Denise Hinzke}
\affiliation{Fachbereich Physik, Universit\"{a}t Konstanz, 78457 Konstanz, Germany}

\author{Tao Liu}
\affiliation{Institute of Physics, Chinese Academy of Sciences, Beijing 100190, China}

\author{Yichuan Wang}
\affiliation{Department of Physics, University of York, York, YO10 5DD, UK}
\affiliation{York-Nanjing International Joint Center in Spintronics, School of Electronic Science and Engineering, Nanjing University, Nanjing, 210093, China}

\author{Tuyuan Cheng}
\affiliation{Spintronics and Nanodevice Laboratory, Department of Electronics, University of York, York YO10 5DD, UK}

\author{Jing Wu}
\email[Email:~]{jing.wu@york.ac.uk}
\affiliation{Department of Physics, University of York, York, YO10 5DD, UK}
\affiliation{York-Nanjing International Joint Center in Spintronics, School of Electronic Science and Engineering, Nanjing University, Nanjing, 210093, China}

\author{Thomas A. Ostler}
\affiliation{Faculty of Arts, Computing, Engineering and Sciences, Sheffield Hallam University, Sheffield, S1 1WB, UK}
\affiliation{D\'{e}partement de Physique, L'Universit\'{e} de Li\`{e}ge, B-4000 Li\`{e}ge, Belgium}

\author{Jianwang Cai}
\email[Email:~]{jwcai@iphy.ac.cn}
\affiliation{Institute of Physics, Chinese Academy of Sciences, Beijing 100190, China}

\author{Ulrich Nowak}
\affiliation{Fachbereich Physik, Universit\"{a}t Konstanz, 78457 Konstanz, Germany}

\author{Roy W. Chantrell}
\affiliation{Department of Physics, University of York, York, YO10 5DD, UK}

\author{Ya Zhai}
\affiliation{Spintronics and Nanodevice Laboratory, Department of Electronics, University of York, York YO10 5DD, UK}
\affiliation{Department of Physics, Southeast University, Nanjing, 210096, China}

\author{Yongbing Xu}
\email[Email:~]{yongbing.xu@york.ac.uk}
\affiliation{York-Nanjing International Joint Center in Spintronics, School of Electronic Science and Engineering, Nanjing University, Nanjing, 210093, China}
\affiliation{Spintronics and Nanodevice Laboratory, Department of Electronics, University of York, York YO10 5DD, UK}


\date{\today}

\begin{abstract}
Using the time-resolved magneto-optical Kerr effect (TR-MOKE) method, helicity-dependent all-optical magnetization switching (HD-AOS) is observed in ferrimagnetic TbFeCo films. Our results reveal the individual roles of the thermal and nonthermal effects after single circularly polarized laser pulse. The evolution of this ultrafast switching occurs over different time scales and a defined magnetization reversal time of 460 fs is shown - the fastest ever observed. Micromagnetic simulations based on a single macro-spin model, taking into account both heating and the inverse Faraday effect, are performed that reproduce HD-AOS demonstrating a linear path for magnetization reversal. 

\end{abstract}


\maketitle
Since the demonstration of magnetization reversal by a single femtosecond laser pulse in 2007~\cite{1}, the field of all-optical switching (AOS) has been extensively studied both theoretically and experimentally. AOS in the ferrimagnetic alloy, GdFeCo (the initially investigated material for AOS), has been shown reverse through a purely thermal effect~\cite{2,3,4,5} where the dynamics proceed via a transient ferromagnetic-like state~\cite{6,52}. Very recently, ultrafast electronic heat currents have been shown experimentally to be sufficient to switch the magnetization in this same material~\cite{7,8}, which provides further evidence of the thermal origins of AOS in GdFeCo~\cite{37}. Consequently, AOS in GdFeCo is almost independent of the laser helicity of the laser pulse, which is named helicity-independent AOS (HI-AOS).

On the other hand, there are many examples of AOS observed in other materials, that are strongly helicity dependent, e.g. ferromagnetic Co/Pt multilayers~\cite{9}, FePt nanoparticles~\cite{39}, synthetic ferrimagnetic heterostructures~\cite{10} and Tb-based ferrimagnets~\cite{11,12,13}. For these materials, there is a one-to-one correspondence of the helicity of the laser light controls and the magnetization orientation, deemed helicity dependent AOS (HD-AOS). A dependence on helicity was observed in GdFeCo for single pulses applied to the alloy for a narrow range of fluence~\cite{14}, which was quantitatively explained as arising from magnetic circular dichroism (MCD)~\cite{15}. Besides the purely thermal effect and MCD~\cite{42}, other mechanisms have been proposed to explain the observed AOS, e.g. inverse Faraday effect (IFE)~\cite{1,27,45,46}, stimulated Raman scattering~\cite{47,48}, sublattice exchange relaxation~\cite{49}, ultrafast exchange scattering~\cite{50}, and optical spin pumping~\cite{53}. However, the underlying physics of HD-AOS in a larger variety of materials is still unclear, especially of the roles of the helicity and thermal effects of the laser pulse. Several experimental criteria and models have been proposed to interpret HD-AOS. A so-called low-remanence criterion was reported whereby HD-AOS is only obtained below a magnetization remanence threshold of $\SI{220}{emu/cm^3}$ for several materials~\cite{12}. Recently, a domain size criterion for the observation of HD-AOS has been proposed, whereby the laser spot size should be smaller than the equilibrium size of magnetic domains forming during the cooling process after laser irradiation~\cite{16}. Meanwhile, using a time-dependent anomalous Hall effect technique, HD-AOS has been demonstrated to consist of a steplike helicity-independent multiple-domain formation followed by a helicity-dependent remagnetization~\cite{17}. There have been several models of optical switching presented in the literature, as well as differing measurements with different conclusions as to the importance of the thermal or nonthermal effects~\cite{4,14,22,42,43,*44}. In this context, one intuitive question is: can the contributions of both thermal and nonthermal effects be quantified simultaneously during a single circularly polarized laser pulse? The ultrafast laser-induced demagnetization is well-known to have a thermal aspect~\cite{18}, however, there will inevitably be some contribution from both thermal and nonthermal effects during one single laser pulse. However, in all the Kerr or Faraday image detections, it is impossible to measure both of these two effects because only the final static magnetization states are observed - one requires access to temporal information.    

To explore the roles of the thermal and non-thermal effects in HD-AOS, and the time scales in this process, we used laser pump-probe technique, also known as time-resolved magneto-Kerr effect measurement~\cite{19} (details in Supplemental Materials), to measure the transient magnetization change after a single laser pulse acting on TbFeCo. The transient reflectivity change is simultaneously monitored. TbFeCo is a similar ferrimagnet compared to GdFeCo as the Tb sublattice is antiferromagnetically coupled with the FeCo sublattice~\cite{20,21,22}, forming a ferrimagnetic structure. However, because of the large difference between the spin-orbit coupling of Tb and Gd~\cite{23}, Gd- and Tb- based alloys show different spin dynamics as well as distinct switching mechanisms~\cite{24,25}.

\begin{figure}
\includegraphics[width=3.2in]{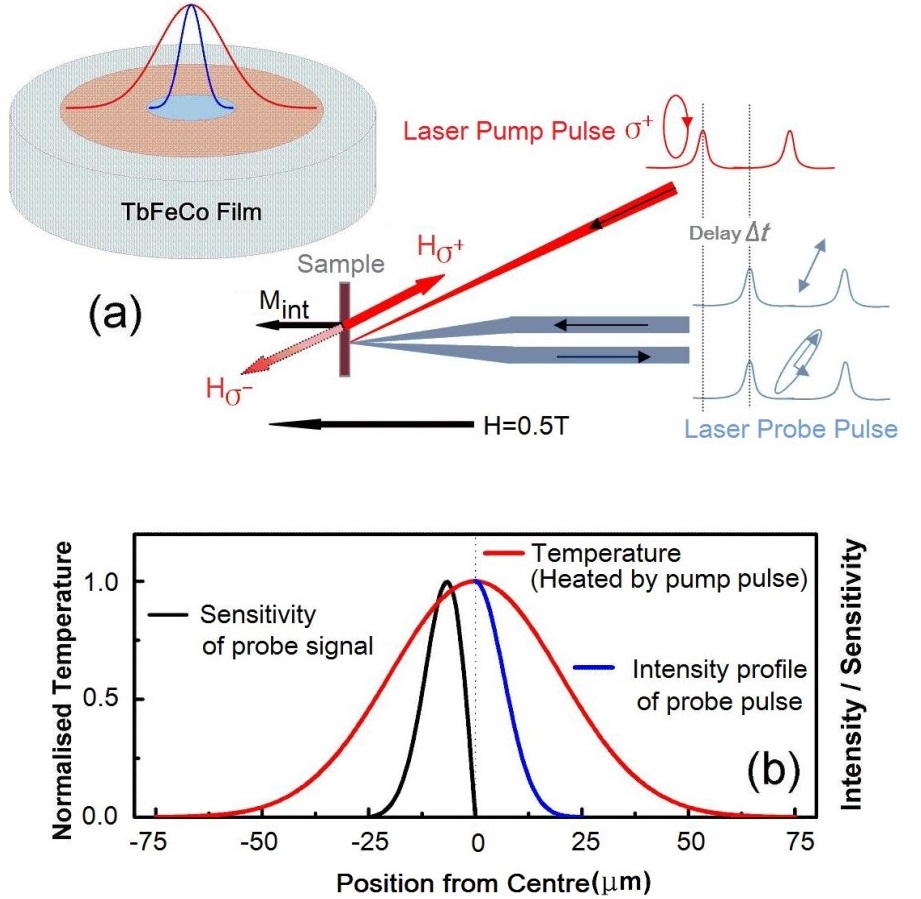}
\caption{\label{fig1}(a) a schematic diagram of the experimental set-up with a bias field $H=\SI{0.5}{\tesla}$. $H_{\sigma^+}$ represents the effective field of the pump pulse with $\sigma^+$ polarization (red line) due to the IFE. (b) the normalized radial sensitivity of Kerr rotation (only left half shown for clarity) and temperature distribution across the pump spot together with the intensity profile of the probe spot (only right half shown for clarity).} 
\end{figure}

\begin{figure}
	\includegraphics[width=3.375in]{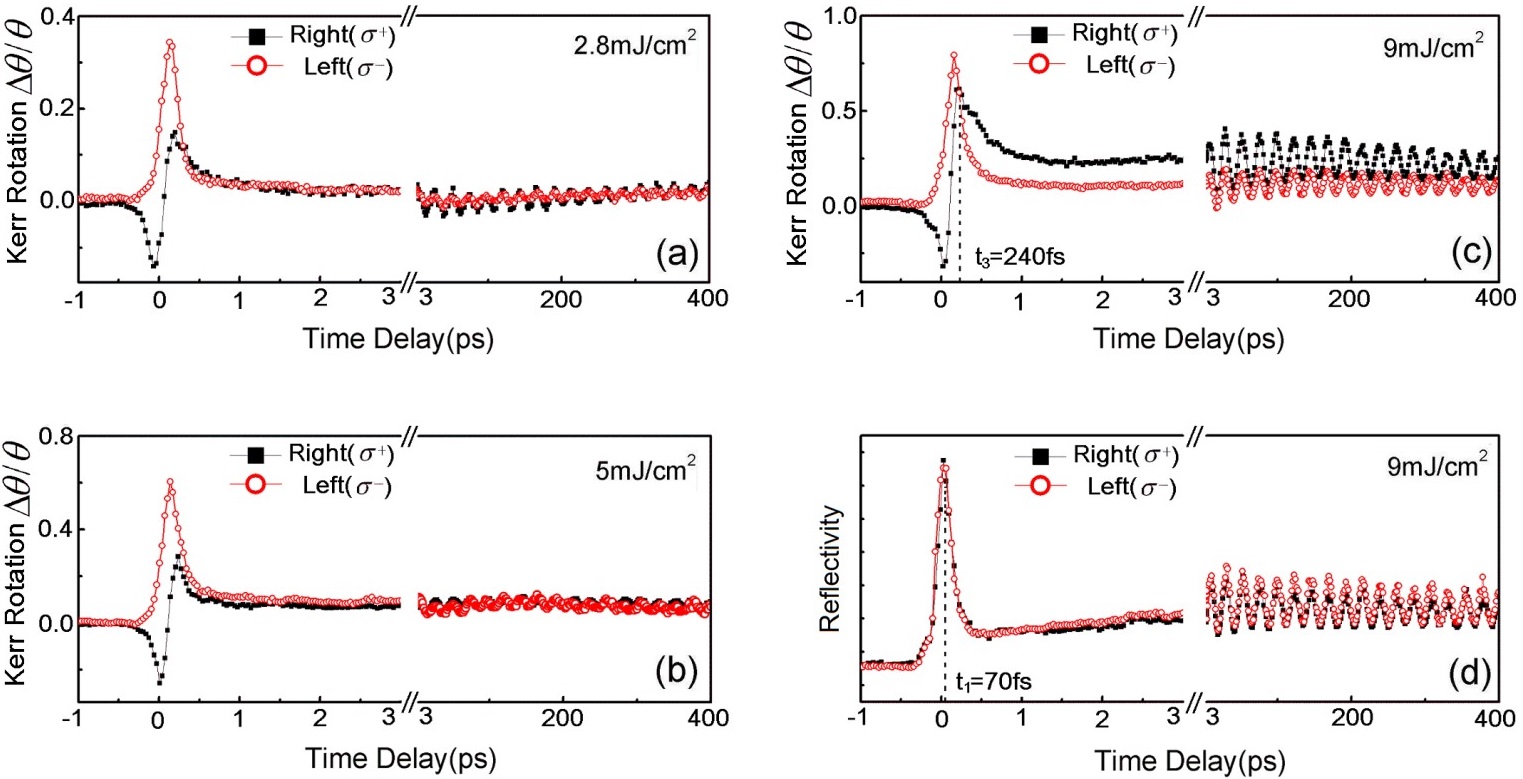}
	\caption{\label{fig2}(a) and (b) show the time domain Kerr rotation taken under a pump fluence of $2.8$ and $\SI[per-mode=symbol]{5}{\milli\joule\per\square\centi\metre}$, respectively. (c) presents the time domain Kerr rotation obtained under pump beam fluence $\SI[per-mode=symbol]{9}{\milli\joule\per\square\centi\metre}$. At about $\SI{240}{\femto\second}$ time delay, the curve excited by pump pulses of $\sigma^+$ polarization (black solid squares) starts to switch further away from the initial magnetization direction compared with the curve excited by $\sigma^-$ polarized  (red hollow dots) pump pulses. (d) shows the time domain reflectivity data at $\SI[per-mode=symbol]{9}{\milli\joule\per\square\centi\metre}$ for both $\sigma^+$ and $\sigma^-$ polarization. The two curves overlap with the peak at $t_{1}=\SI{70}{\femto\second}$, indicating the maximal electron temperature. } 
\end{figure}

In order to separate thermal and nonthermal contributions, time domain measurements are performed, varying the laser pump fluence and helicity, whilst keeping the direction of the external magnetic field fixed in the direction almost parallel to the direction of the induced magnetization due to the $\sigma^-$ helicity pulses (and nearly anti-parallel in the $\sigma^+$ case). The transient Kerr rotation obtained under different laser fluences with different laser helicities are shown in Fig.~\ref{fig2}(a-c). Between the two lower laser fluences (2.8 and $\SI[per-mode=symbol]{5}{\milli\joule\per\square\centi\metre}$), the dynamic responses are very similar except that the amplitude is increased with the laser fluence. The two curves taken with different laser helicity converge after around $\SI{240}{\femto\second}$ time delay, suggesting that only thermal effects exist for these laser fluences because the thermal effects are insensitive to the laser helicity while the nonthermal effect is~\cite{43}. The peaks around zero delay are the so-called specular inverse Faraday effect (SIFE) and specular optical Kerr effect (SOKE) contributions~\cite{26}, as detailed in Fig. S1 in the Supplemental Materials. However, as the laser fluence is increased to $\SI[per-mode=symbol]{9}{\milli\joule\per\square\centi\metre}$, the two curves taken with different laser helicity no longer converge. The curve excited by laser pulses of $\sigma^+$ polarization (a helicity that induces an effective field opposite to the external magnetic field) switches further away from the initial magnetization direction compared to the curve excited by $\sigma^-$ polarised laser pulses. This extra switching starts at around $t_{3}=\SI{240}{\femto\second}$, indicating the onset of the nonthermal effect. The time evolution of the reflectivity has also been investigated indicating a peak electron temperature at approximately $t_{1}=\SI{70}{\femto\second}$. There is no obvious laser helicity dependence in the reflectivity which can be seen from the data taken at $\SI[per-mode=symbol]{9}{\milli\joule\per\square\centi\metre}$ as shown in Fig.~\ref{fig2}(d). In this case, the absorptions of light are at the same level as well as the electron temperature profiles which means there is no significant MCD effect. The oscillations with a high frequency of $\SI{42}{GHz}$ shown in both the transient Kerr rotation and reflectivity data have no magnetic field dependence. Therefore, it may be originated from a laser-induced strain-wave in the amorphous films (details in Supplemental Materials).

\begin{figure}[t]
	\includegraphics[width=3.375in]{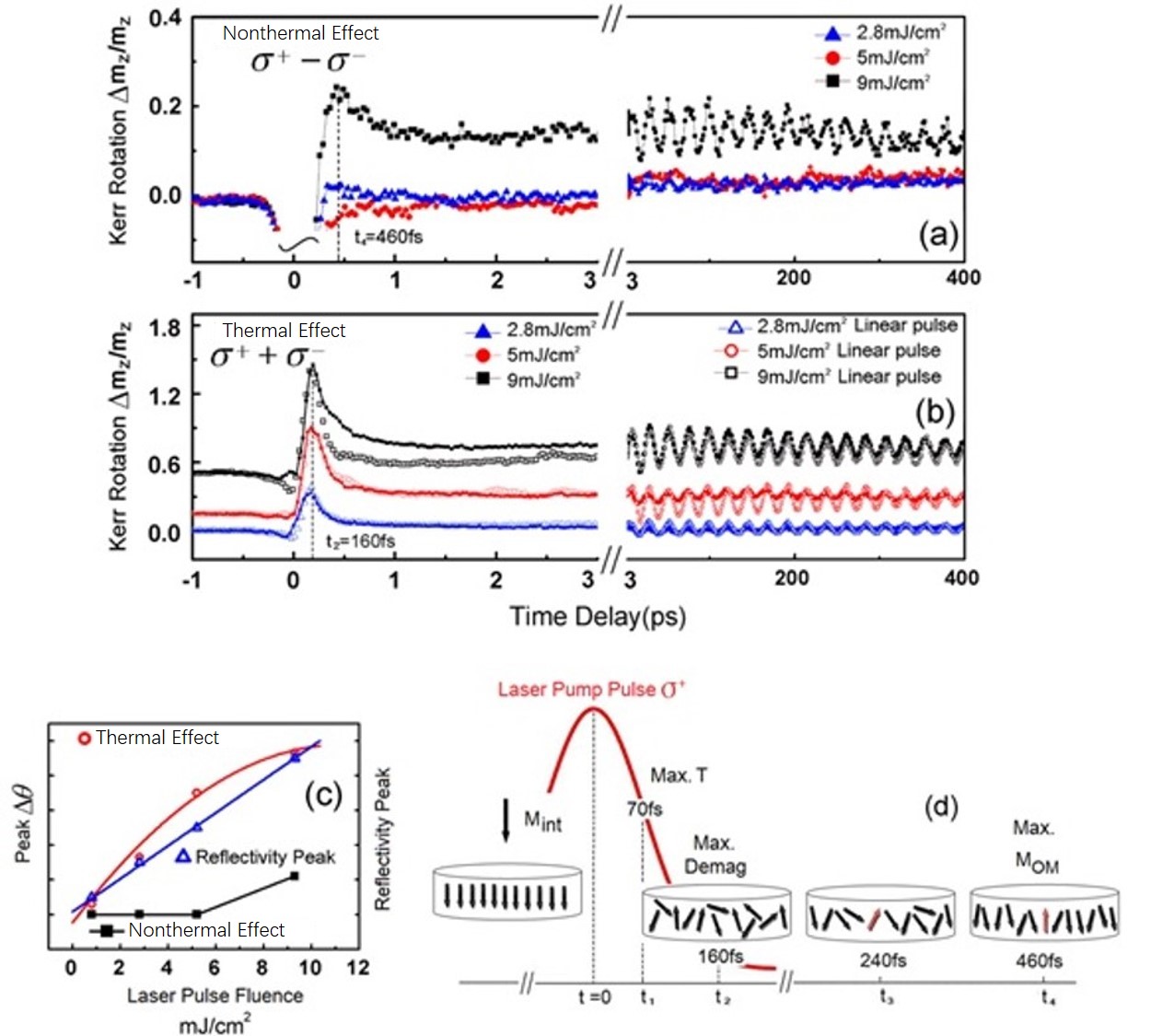}
	\caption{\label{fig3}(a) shows the difference between the $\sigma^+$ and $\sigma^-$ pump pulses as a function of time. (b) The three solid curves show the sum of the $\sigma^+$ and $\sigma^-$ pump pulses with time. The hollowed curves are time domain responses excited by the linearly polarized laser pulse at the same pump fluences. (c) The peak amplitude of the thermal effect (red circles), of the reflectivity (blue triangles), and of the nonthermal effect (black squares) at a delay time of $\SI{460}{\femto\second}$ as a function of the pump fluence. (d) Shows a schematic diagram of the ultrafast process induced at $\SI[per-mode=symbol]{9}{\milli\joule\per\square\centi\metre}$ pump fluence.} 
\end{figure}

The thermal and nonthermal effects on the magnetization can be separated by analysing respectively the sum and difference of the experimental data under different laser helicities. Therefore, the data sets in Fig.~\ref{fig2}(a-c) have been analysed accordingly and are presented in Fig.~\ref{fig3}(a-b). The difference data in Fig.~\ref{fig3}(a) shows the time evolution of the nonthermal effect. For the two cases with lower laser fluence, the time evolution of the two difference data overlaps and goes back to its original state immediately after the SIFE/SOKE peak, giving no indication of any nonthermal effect. As the pump fluence is increased to $\SI[per-mode=symbol]{9}{\milli\joule\per\square\centi\metre}$, the difference signal does not return to the original state immediately. Instead it keeps increasing to its maximum magnitude at around $t_{4}=\SI{460}{\femto\second}$ time delay showing that the magnetization has partially switched in some regions of the irradiated area to a different magnetization state. This demonstrates unambiguously a helicity-dependent switching in TbFeCo triggered at close to $t_{3}=\SI{240}{\femto\second}$ and magnetization re-orientation at approximately $t_{4}=\SI{460}{\femto\second}$ after circularly polarized laser excitation.  

Fig.~\ref{fig3}(b) presents the time evolution of the directly measured heat-driven dynamics excited by a linearly polarized laser of the same energy, along with the data obtained by taking the sum of the $\sigma^+$ and $\sigma^-$ cases for three different laser fluences. All three pairs of time domain Kerr rotation data reach maxima around $t_{2}=\SI{160}{\femto\second}$, indicating the time scale of the quenching of the magnetic order. Two pairs of time domain data taken at lower laser fluence overlap with each other extremely well since the SIFE/SOKE changes phase between $\sigma^+$ and $\sigma^-$ helicity and are thus cancelled out by the sum operation. The pair taken at $\SI[per-mode=symbol]{9}{\milli\joule\per\square\centi\metre}$ start to diverge from each other immediately after the maximum demagnetization with the sum data deviating further from the initial magnetization state, indicating the onset of the helicity-dependent switching excited by $\sigma^+$ pump pulses, which are more profound than those excited by the $\sigma^-$. This is expected, since the helicity-dependent switching induced by two different laser helicities are different in phase as well as in magnitude, depending on the instantaneous magnetization state, and also supported by our theoretical calculations shown below. The peak amplitude of the thermal and reflectivity data is plotted as a function of the pump laser fluence in Fig.~\ref{fig3}(c) together with the amplitude of the nonthermal data at $\SI{460}{\femto\second}$ time delay. Fig.~\ref{fig3}(c) shows that the electron temperature is proportional to the laser fluence; the sample is nearly totally demagnetized at $\SI[per-mode=symbol]{9}{\milli\joule\per\square\centi\metre}$ which is consistent with the condition required for helicity-dependent switching~\cite{17}; there is no sign of helicity-dependent switching for the data taken at lower pump fluence. Note that $\SI[per-mode=symbol]{9}{\milli\joule\per\square\centi\metre}$ is the highest pump fluence, which can be applied without damaging the sample surface, and the helicity-dependent switching is only observed at this highest pump fluence. The whole ultrafast process induced at a pump fluence of  $\SI[per-mode=symbol]{9}{\milli\joule\per\square\centi\metre}$ is schematically summarized in Fig.~\ref{fig3}(d).  The electron temperature reaches its maximum at $\SI{70}{\femto\second}$ time delay and the magnetic order is largely quenched by $\SI{160}{\femto\second}$. The onset of helicity-dependent switching takes place within $\SI{240}{\femto\second}$ and a new magnetization direction is defined by $\SI{460}{\femto\second}$.

\begin{figure}[t]
	\includegraphics[width=3.375in]{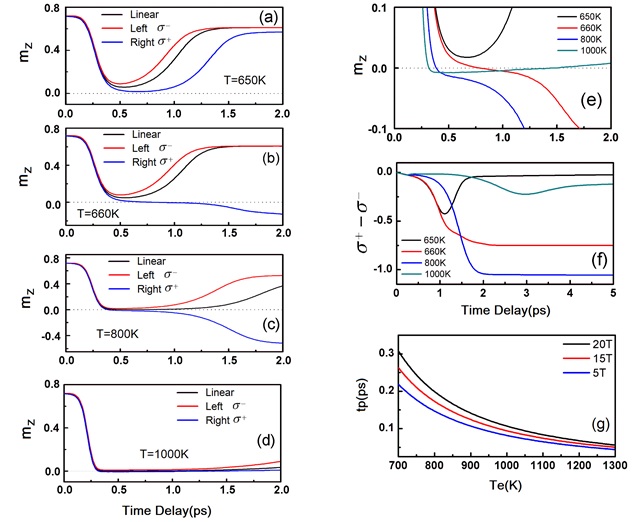}
	\caption{\label{fig4}(a-d) Simulated magnetic response for increasing peak electron temperatures showing the onset of magnetization reversal. Reversal occurs at around $\SI{660}{\kelvin}$, where the linear reversal mechanism sets in. At higher temperatures (d) the magnetization is destroyed after reversal. (e) magnetic response for different values of the peak electron temperature.  (f) shows the difference in the response to the difference helicities. Optically induced reversal is demonstrated at 660 and $\SI{800}{\kelvin}$. The onset of reversal depends critically on the onset of linear reversal. The response time calculated from Eq. S2 demonstrates that the predicted reversal takes place on the sub-$\mathrm{picosecond}$ tiemscale, consistent with the experimental results. (g) Minimal field and temperature pulse time needed to trigger a magnetization reversal taken from Eq. S2.} 
\end{figure}

To understand the observed time domain results of HD-AOS, two main effects are considered, namely the MCD~\cite{15} and the IFE~\cite{27}. MCD leads to a different absorption of the two circular helicities in the different domains and it is excluded because from the transient reflectivity curves, no difference is observed with respect to the laser helicity. In Ref.~\onlinecite{39}, the magnetization induced through the IFE effect was directly calculated for the case of FePt with ab-initio methods~\cite{40}. In our simulations, due to a lack of ab-initio calculations for the considered TbFeCo alloy, this temporal change of the magnetization caused by the IFE is assumed to be due to an effective magnetic field~\cite{14,34}. Our simulations are based on a single macro-spin model whereby we solve the Landau-Lifshitz-Bloch (LLB) equation numerically~\cite{28,30,31,38,41}. The LLB equation takes into account transient changes in the length of the magnetization required to describe the heating from the laser pulse. All our methods are described in detail in Ref.~\onlinecite{31} and also summarized in the Supplemental Materials. The results of these simulations are shown in Figs.~\ref{fig4}(a-d) for different peak electron temperatures $T_{\rm{e}}$, corresponding to different laser fluences, as summarised in Fig.~\ref{fig4}(e-f). The figure focuses on the change in the reduced magnetization ($M/M_{\rm{s}}$) along the easy-axis at short time-scales. Starting at room temperature, the reduced magnetization at equilibrium is around 0.8. Complete demagnetization can be achieved within $\SI{300}{\femto\second}$ and magnetization reversal can be triggered on the sub-$\mathrm{picosecond}$ time-scale for higher $T_{\rm{e}}$. The theoretical model reproduces the sub-$\mathrm{picosecond}$ reversal observed experimentally and confirms the above interpretation of the experimental data. Above all, the reversal occurs only above a critical temperature corresponding to that of the linear reversal model; reversal on this timescale cannot occur via precessional mechanisms, which occur on the nanosecond timescale. Therefore, the peak electron temperature plays a significant role in HD-AOS. In Ref.~\onlinecite{32} an analytical formula was derived for the minimal pulse time (in terms of a rectangular field and temperature pulse), which is needed to switch the sign of the magnetization (see Eq. S2 in the Supplemental Materials). It is illustrated in Fig.~\ref{fig3}(g). We noticed that in the simulations the switching times are slightly larger than with the analytical formula. This is due to the fact that for the analytical formula a rectangular temperature and field pulse is assumed, while in the simulations more realistic profiles are calculated. We also note that the simulations further predict a rapid increase of the magnetization in a negative sense after reversal, whereas the experimental data indicate that the magnetization recovers towards the original value. We attribute this to the simplified nature of the calculations, which are based on a single spin, whereas the experimental sample has a large-scale domain structure, though quantitative agreement is not the aim here. While the reversal of the magnetization via the linear reversal mechanism is unlikely to be affected by the domain structure, it is reasonable to expect that the magnetization measured by the probe beam after the pulse cannot be simulated within the current single spin model. Furthermore, multi-macrospin calculations would most likely still not be comparable with experimental measurements as the size of the probe beam is still many micrometres and likely beyond the size of this type of simulation. It should also be noticed that, compared to the current single macrospin simulations leading to a linear reversal mechanism, an atomistic spins approach would possibly give a different picture, as there would be more degrees of freedom for the atomic spins to relax.

In summary, the HD-AOS is unambiguously demonstrated in a TbFeCo film by one single circularly polarized laser pulse. The thermal and nonthermal effects are seen to have different time scales, respectively. High pump fluences are required to observe laser helicity effects, which is consistent with other reported works~\cite{16,17}. Note that the effect of heat accumulation is not excluded in our measurements, but the $\SI{1}{\kilo\hertz}$ laser repetition rate is much lower than the repetition rate used in Ref.~\onlinecite{12} which shows no significant accumulative heat. Besides, the relaxation time of transient reflectivity response is quite small in our measurements, so the effect of accumulative heat should not play a role. The interplay between laser heating and helicity is stimulated by a single laser pulse. The whole process of the magnetization switching contains four periodes; peak electron temperature achieved; the system becomes fully demagnetized, magnetization switching is triggered; and a new magnetization direction is defined. Furthermore, from our measurements we can see that, on the sub-$\mathrm{picosecond}$ time-scales, there is a magnetization switching time within $\SI{460}{\femto\second}$ - the fastest among the reported times in the literature~\cite{14,33,34, 35}. Very recently, a theoretical study by means of first-principles and model simulation predicts a magnetization switching time of  $\SI{218}{\femto\second}\sim\SI{609}{\femto\second}$~\cite{54}, which is in good agreement with our findings. This sub-$\mathrm{picosecond}$ switching is reproduced using a single macro-spin model based on the stochastic Landau-Lifschitz-Bloch equation, confirming the linear reversal mechanism without spin precession in all-optically induced magnetization switching in TbFeCo. Also, the simulations suggest that heating the electron system to a critical temperature may play an important role in this kind of magnetization reversal. Above all, the finding of ultrafast helicity-dependent all-optical magnetization switching in a high anisotropy system triggered by a single laser pulse brings all-optical magnetic recording a major step closer to high data rate and high data density applications.

See Supplementary Material for details of sample preparation, polar TR-MOKE setup, probe sensitivity, and theoretical modelling. The SIFE/SOKE contribution and the strain waves are also presented.

This work was supported in part by the National Basic Research Program of China (No. 2014CB921101), National Key Research and Development Program of China (No. 2016YFA0300803), the National Natural Science Foundation of China (No. 61427812, 11574137, 11774160),  Jiangsu Shuangchuang Programme and the Natural Science Foundation of Jiangsu Province of China (No. BK20140054 ). T. A. Ostler gratefully acknowledges the support of the Marie Curie incoming BeIPD-COFUND fellowship program at the University of Li\`{e}ge.

%

\end{document}



\title{Supplemental Materials on ``Roles of heating and helicity in ultrafast all-optical magnetization switching in TbFeCo''}


\author{Xianyang Lu}
\affiliation{Department of Physics, University of York, York, YO10 5DD, UK}
\affiliation{York-Nanjing International Joint Center in Spintronics, School of Electronic Science and Engineering, Nanjing University, Nanjing, 210093, China}

\author{Xiao Zou}
\affiliation{Department of Physics, University of York, York, YO10 5DD, UK}

\author{Denise Hinzke}
\affiliation{Fachbereich Physik, Universit\"{a}t Konstanz, 78457 Konstanz, Germany}

\author{Tao Liu}
\affiliation{Institute of Physics, Chinese Academy of Sciences, Beijing 100190, China}

\author{Yichuan Wang}
\affiliation{Department of Physics, University of York, York, YO10 5DD, UK}
\affiliation{York-Nanjing International Joint Center in Spintronics, School of Electronic Science and Engineering, Nanjing University, Nanjing, 210093, China}

\author{Tuyuan Cheng}
\affiliation{Spintronics and Nanodevice Laboratory, Department of Electronics, University of York, York YO10 5DD, UK}

\author{Jing Wu}
\email[Email:~]{jing.wu@york.ac.uk}
\affiliation{Department of Physics, University of York, York, YO10 5DD, UK}
\affiliation{York-Nanjing International Joint Center in Spintronics, School of Electronic Science and Engineering, Nanjing University, Nanjing, 210093, China}

\author{Thomas A. Ostler}
\affiliation{Faculty of Arts, Computing, Engineering and Sciences, Sheffield Hallam University, Sheffield, S1 1WB, UK}
\affiliation{D\'{e}partement de Physique, L'Universit\'{e} de Li\`{e}ge, B-4000 Li\`{e}ge, Belgium}

\author{Jianwang Cai}
\email[Email:~]{jwcai@iphy.ac.cn}
\affiliation{Institute of Physics, Chinese Academy of Sciences, Beijing 100190, China}

\author{Ulrich Nowak}
\affiliation{Fachbereich Physik, Universit\"{a}t Konstanz, 78457 Konstanz, Germany}

\author{Roy W. Chantrell}
\affiliation{Department of Physics, University of York, York, YO10 5DD, UK}

\author{Ya Zhai}
\affiliation{Spintronics and Nanodevice Laboratory, Department of Electronics, University of York, York YO10 5DD, UK}
\affiliation{Department of Physics, Southeast University, Nanjing, 210096, China}

\author{Yongbing Xu}
\email[Email:~]{yongbing.xu@york.ac.uk}
\affiliation{York-Nanjing International Joint Center in Spintronics, School of Electronic Science and Engineering, Nanjing University, Nanjing, 210093, China}
\affiliation{Spintronics and Nanodevice Laboratory, Department of Electronics, University of York, York YO10 5DD, UK}


\date{\today}




\maketitle
\section{Sample preparation}
The $\SI{20}{\nm}$ Tb$_{19}$Fe$_{66}$Co$_{15}$ thin film was deposited onto a Corning 7059 glass substrates with a $\SI{20}{\angstrom}$ Ta underlayer at ambient temperature using dc magnetron sputtering. A composite target, made by symmetrical placement of Tb and Co chips on the Fe target, was used to deposit TbFeCo films with a composition determined by inductively coupled plasma-atomic emission spectroscopy (ICP-AES). Sputtering rates for TbFeCo and Ta are about $0.9$ and $\SI{1.1}{\angstrom/s}$, respectively. The TbFeCo film was covered by a $\SI{40}{\angstrom}$ Au layer to protect against oxidation. The TbFeCo film exhibits a strong perpendicular anisotropy with a coercive field of $\SI{0.39}{T}$ measured by vibrating sample magnetometry (VSM) at room temperature (Fig. S1).

\begin{figure}[]
	\includegraphics[width=3.375in]{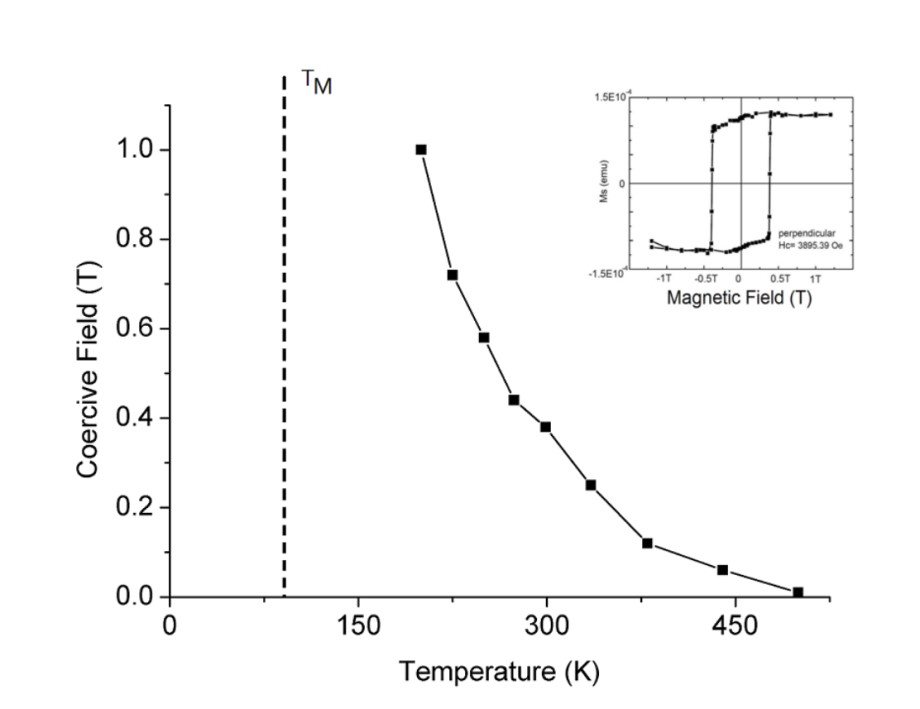}
	\caption{\label{figS1}The compensation temperature $T_{\rm{Mcomp}}$ was estimated from the variationof the coercive fields as a function of the temperature, which is around $\SI{100}{\kelvin}$. The Curie temperature measured is about $\SI{500}{\kelvin}$. The inset picture shows the hysteresis loop of the TbFeCo film measured by VSM at room temperature.} 
\end{figure}

\section{Polar TR-MOKE}
A typical polar TR-MOKE setup is used for all-optical pump-probe measurements. An ultrafast regenerative Ti:sapphire laser system with a pulse width of $\SI{150}{fs}$, central wavelength $\SI{800}{nm}$, and repetition rate of $\SI{1}{kHz}$ was used. The polarization of the $\SI{800}{nm}$ pump beam is varied between linear and circular using a $\lambda/4$ waveplate while the polarization of the $\SI{400}{nm}$ probe beam remains linear. The pump pulses are incident onto the sample at 10 degrees to the sample normal and the probe pulses at normal incidence. Both pump and probe beam are focused onto the same spot on the sample using two lenses with different focal length and positioned with the help of a CCD camera. The spot diameter is around $\SI{150}{\micro\meter}$ for the pump and $\SI{50}{\micro\meter}$ for the probe. The normalized intensity of the pump and probe beams can be described by a Gaussian distribution. Assuming the sensitivity is proportional to the beam intensity and the probing area, the probe sensitivity is calculated as shown in Fig.1(b). The probe is sensitive to a wide range of temperatures and most sensitive to materials in the temperature range from 80 to $\SI{97}{\%}$ of the peak temperature at the centre of the pump/probe. The temporal overlap of the laser pump pulse and the laser probe pulse is found to be about $\SI{250}{\femto\second}$ FWHM. The rotation of the polarization and change of the reflectivity of the reflected probe beam is detected. To maintain the same initial magnetic state, a magnetic field $H_{ext}=\SI{0.5}{\tesla}$, which is $\SI{0.1}{\tesla}$ higher than the sample coercive field at room temperature, is applied perpendicular to the sample plane. 

\section{Probe sensitivity}
In order to quantitatively analyse the relationship between the detected signals and the probe area, the induced temperature is assumed to be proportional to the pump intensity, which can be described by a Gaussian distribution (red curve in Fig. 1(b)), and the change of the magnetization (thermally) is proportional to the temperature. The intensity profile of the probe pulse (right half shown in blue in Fig. 1(b)) is also described by a Gaussian distribution with a one-third spot diameter of the pump. The probe sensitivity is then assumed to be proportional to the product of the beam intensity $I_{\rm{probe}}(r)={\rm{A}} e^{-{\rm{\alpha}} r^{2}}$, the change of the magnetization $\Delta M(T,r)={\rm{B}} e^{-{\rm{\beta}} r^{2}}$ and the area generating the signal; where $\rm{A}$, $\rm{B}$ are the normalized coefficients, the parameters $\alpha$ and $\beta$ are determined by the pump and probe beam diameters to characterize a Gaussian distribution, $r$ is the distance from the centre of the pump spot. The probe spot has circular symmetry and so the probed area equal to $2\pi r{\rm{d}}r$, both the pump and probe beams have a constant intensity, therefore corresponding to a constant temperature and thus a constant change of magnetization. Therefore, the probe sensitivity can be calculated by
\begin{equation}
\label{eqS2}\Delta M(T,r)\times I_{\rm{probe}}(r)\times 2\pi r{\rm{d}}r = {\rm{C}}re^{-{\rm{D}} r^{2}}{\rm{d}}r.
\end{equation}
It is clear that the probe sensitivity is not the highest at $r=0$ though the beam is brightest here. From the calculation (left half shown in black in Fig. 1(b)), it is indicated that the probe beam detects $\Delta M(T)$ caused by different temperature ranges from $80$ to $97\%$ of the peak temperature at the centre of the pump/probe overlap. 
 
\section{Theoretical modelling}
The treatment of ultrafast dynamics at elevated temperature requires an equation of motion, which allows for transient changes in the length of the magnetization. The LLB equation~\cite{1} encapsulates very well the response of a set of coupled atomic spins subjected to rapidly varying temperature changes, including the reduction of the magnitude of the magnetization~\cite{2}. The equation of motion and all relevant methods are described in detail in Ref.~\onlinecite{6}. The temperature-dependent parameters for the LLB equation, i.e., longitudinal and transverse susceptibilities and the temperature variation of the magnetization, are calculated numerically from an atomistic simulation using the stochastic Landau-Lifshitz-Gilbert equation for each atomic spin~\cite{3}. These numerically gained temperature dependent material parameters as calculated previously for FePt~\cite{3}, are rescaled to reflect a ferromagnetic material with a Curie temperature of $T_{\rm{C}}=\SI{500}{\kelvin}$, a saturation magnetization $M_{s}$ of $\SI{2.5e5}{\ampere\per\metre}$, and a room temperature anisotropy $K$ of $\SI{3.8e5}{\joule\per\cubic\metre}$ (see also Table S1). We consider a single macro-spin with a volume of $(\SI{30}{nm})^3$ and averaged for all simulations over 100 runs. The microscopic damping constant was $\lambda=0.1$. As in the experiment, an additional constant field of $\SI{0.5}{\tesla}$ was applied in magnetization direction. Consistent with previous simulations~\cite{2}, the electron temperature $T_{\rm{e}}$ serves as the temperature of the heat bath for our simulation. This temperature is estimated from a two-temperature model (for more details see Ref.~\onlinecite{6} and references within) assuming a laser pulse of $\SI{170}{\femto\second}$ pulse width. The electronic specific heat is $C_{\rm{e}}(T_{\rm{e}})=\SI[parse-numbers=false]{700\textit{T}_{\rm{e}}}{\joule\per{\kelvin\cubic\metre}}$, the lattice specific heat is $C_{\rm{l}}=\SI{3e6}{\joule\per{\kelvin\cubic\metre}}$ and the electron-phonon coupling constant is $C_{\rm{el}}=\SI{1.7e18}{\joule\per{\second\kelvin\cubic\metre}}$. Here, due to a lack of ab-initio calculation as described in \cite{7,8}, it is assumed that due to the inverse Faraday effect the circularly polarized laser pulse triggers a magnetic field pulse $B_{IFE}$, the height of which is estimated to be $\SI{20}{\tesla}$~\cite{4,9,10}. We assume for its duration time $t_{\rm{B}}=\SI{400}{\femto\second}$. The direction of this field is determined by the helicity of the laser pulse. 

In Ref.~\onlinecite{5} an analytical formula was derived for the minimal pulse time (in terms of a rectangular field and temperature pulse), which is needed to switch the sign of the magnetization. For temperatures above the Curie temperature, Eq. 9 from that paper can be used where the arctan term can be neglected for smaller magnetic fields and higher temperatures, yielding
\begin{equation}
\label{eqS1}t_{\rm{p}}\approx-\frac{3\tilde{\chi}_{\parallel}T_{\rm{C}}}{2\gamma\lambda T_{\rm{e}}}\ln(\tilde{\chi}_{\parallel}B\sqrt{\frac{T_{\rm{e}}-2T_{\rm{C}}/5}{T_{\rm{e}}-T_{\rm{C}}}}).
\end{equation}  
For our numerically determined longitudinal susceptibility, the behaviour of this equation is illustrated in Fig. 4(f).

\begin{figure}
	\includegraphics[width=3.375in]{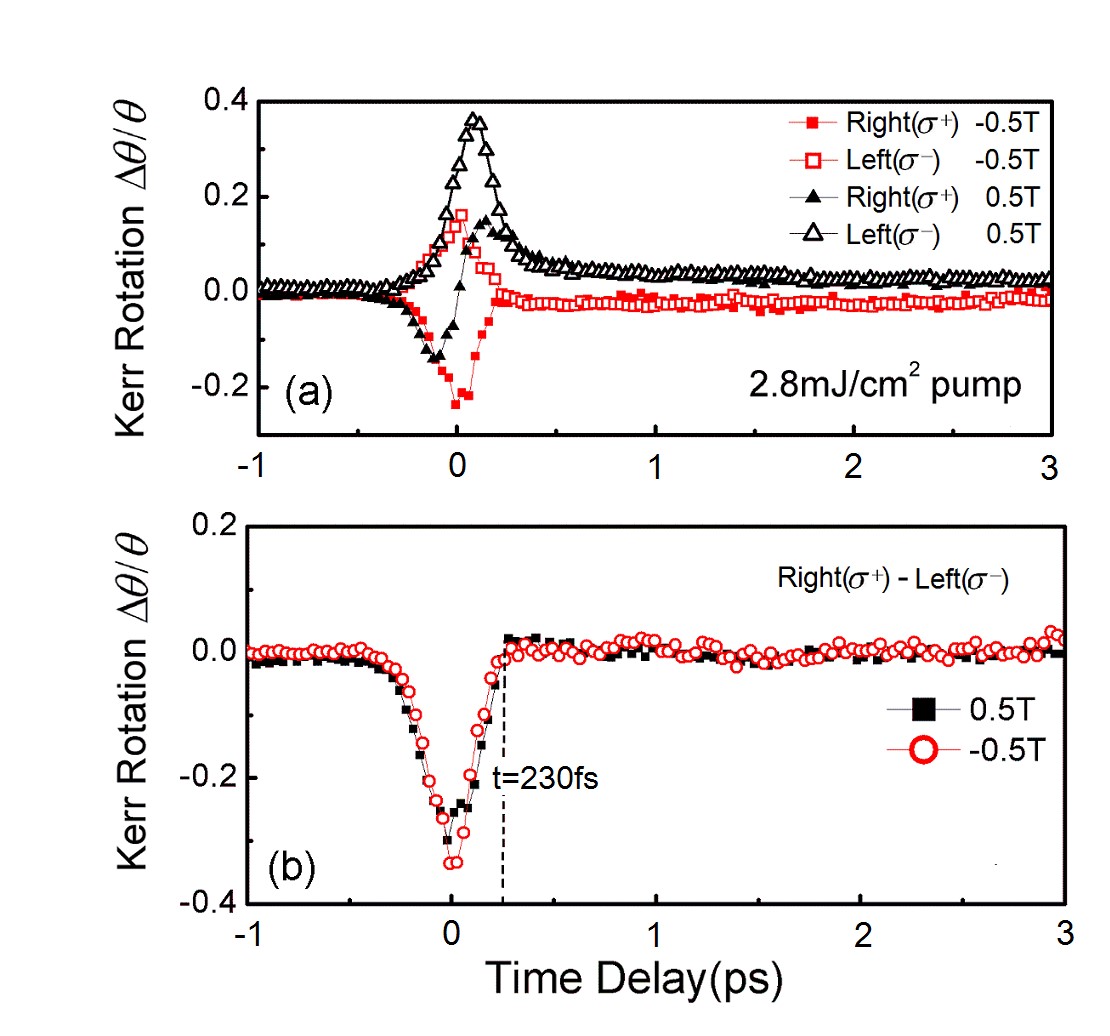}
	\caption{\label{figS2}(a) shows the time domain Kerr rotation data of magnetization dynamics excited by Right($\sigma^+$)/Left($\sigma^-$) circularly polarized pump pulses at pump fluence $\SI{2.8}{\milli\joule\per\square\centi\meter}$ in a TbFeCo thin film under $+/-\SI{0.5}{\tesla}$ external fields, presenting a mixture of both ultrafast demagnetization (heat-driven effect) and optical (including opto-magnetic and optically-induced birefringence) signal. (b) presents the difference between the Right($\sigma^+$) and the Left($\sigma^-$) at the same magnetic field for two field directions ($+/-\SI{0.5}{\tesla}$), indicating the birefringence effect is insensitive to magnetization. The zero time delay is set at the maximum of the difference data. The symmetric peak with $\SI{250}{fs}$ FWHM at the zero time delay demonstrates that the optical effects have only occurred during this temporal window and vanish around $\SI{230}{fs}$.} 
\end{figure}

\begin{table}[hbt]
	\caption{\label{tab:S1}Model parameters used in the LLB simulations compared to those of Ref.~\onlinecite{10}.}
	\begin{ruledtabular}
		\begin{tabular}{llllll}
			& $T_{C} (\SI{}{K})$  & $K (\SI{}{\joule\per\cubic\meter})$ & $B_{IFE} (\SI{}{T})$ & $t_{B} (\SI{}{fs})$ & $M_{s} (\SI{}{A\per\meter})$\\
			GdFeCo\cite{10} & 500 & $\SI{6.05e5}{}$ & 20 & 240 & $\SI{1.3e6}{}$\\
			TbFeCo          & 500 & $\SI{3.80e5}{}$ & 20 & 400 & $\SI{2.5e5}{}$\\
		\end{tabular}
	\end{ruledtabular}
\end{table}

\section{SIFE/SOKE contribution}
To understand the behaviour within the first two hundred femtoseconds, two more scans in addition to those shown in Fig. 2(a) have been taken at the same pump laser fluence $\SI{2.8}{\milli\joule\per\square\centi\metre}$ with the $\SI{0.5}{\tesla}$ bias field reversed and close to parallel to the $\sigma^+$ helicity. The complete data set taken under different pump beam helicities and different directions of bias magnetic fields is shown in Fig.~\ref{figS2}(a). Within the first two hundred femtosecond time delay, the dynamic responses are very different between the four different combinations of pump helicity and magnetic field direction, presenting a mixture of both ultrafast demagnetization (thermal effect) and the SIFE/SOKE contribution. Around $\SI{230}{fs}$ time delay, the two curves taken under the same magnetic field direction converge, indicating that the effect due to different pump helicity is vanishing, leaving only the thermal effect corresponding to this pump laser fluence. For a fixed magnetic field direction, the thermal effect is the same no matter what the polarization of the pump beam, while the SIFE/SOKE change phase. Therefore the difference between the data of right and left helicity at the same magnetic field is sensitive to the SIFE/SOKE, as shown in Fig.~\ref{figS2}(b). The symmetric peak with $\SI{250}{fs}$ FWHM at the zero delay in Fig.~\ref{figS2}(b) demonstrates that the SIFE/SOKE have only occurred during this temporal window. The $\SI{250}{fs}$ FWHM is the width of the cross-correlation between the pump and probe beam with $\SI{170}{fs}$ pulse width, which is a reasonable value after passage through the external optics. The symmetric peak suggests that the SIFE/SOKE exist only during the presence of the pump pulse and vanishes completely at $\SI{230}{fs}$ time delay. Both the opto-magnetic effect and the SIFE/SOKE depend on pump helicity. However, the induced birefringence only persists during the presence of the pump pulses, while the opto-magnetic effect on the sample magnetization will last after the pump pulses disappear. We surmise therefore, that any helicity dependence in the time-domain Kerr rotation data after the first $\SI{230}{fs}$ time delay is due to the opto-magnetic effect. On the other hand, reversal of the magnetic field under the same pump helicity would not invert the curves symmetrically as shown in Fig. 1(a). This asymmetry is attributed to the contribution which arise from the pump-induced change of optical anisotropies~\cite{11} or from the transient reflectivity change~\cite{12}. Nevertheless, this contribution is magnetization-independent which would not affect our experiments.

\section{STRAIN WAVES}
No precessional motion is observable in the time evolution data of Fig. 3(a). This may be due to the strong ringing at $\SI{42}{\giga\hertz}$ frequency, which is present in both the Kerr rotation and reflectivity data. We found that the period of this ringing has no magnetic field dependence and no film thickness dependence. The origin of this ringing may be from the strain waves excited and manipulated by femtosecond laser pulses. The reflectivity oscillates because the optical constants of the thin film are changed by the propagating strain waves. The period t of the oscillations is $t=\lambda/2nv$, where n is the real part of refractive index (2.3 for TbFeCo film) and $\lambda$ is wavelength ($\SI{400}{nm}$) for the probe. This gives a reasonable sound velocity of $\SI{2800}{\meter\per\second}$ for TbFeCo, which is very close to the sound speed $\SI{2620}{\meter\per\second}$ of Terbium bulk material and smaller than the sound velocities of Iron ($\SI{5100}{\meter\per\second}$) and Cobalt ($\SI{4700}{\meter\per\second}$).